\documentclass[12pt]{article}

\usepackage{subfigure}
\usepackage{graphicx}
\usepackage{floatflt}
\usepackage{wrapfig}

\begin{document}

\title{ \textsc{frank-condon principle\\ and adjustment of optical waveguides\\ with nonhomogeneous refractive indices}}
\author{\bf Vladimir I. Man'ko$^{1}$, Leonid D. Mikheev$^{1}$, and Alexandr Sergeevich$^{2*}$}
\date{}
\maketitle

\centerline{$^{1}$\textit{P. N. Lebedev Physical Institute, Russian
Academy of Sciences }}

\centerline{\textit{Leninskii Prospect 53, Moscow 119991, Russia}}
\vspace{1mm} \centerline{$^{2}$\textit{School of Physics, The
University of Sydney}} \centerline{\textit{New South Wales 2006,
Australia}}

\vspace{2mm} \centerline{*Corresponding author, e-mail: $\;\;\;$
a.sergeevich@physics.usyd.edu.au}\vspace{1mm} \centerline{e-mails:
manko@sci.lebedev.ru$\;\;$ mikheev@sci.lebedev.ru} \vspace{2mm}
\def\thesection{\arabic{section}.}
\makeatletter
\renewcommand\@biblabel[1]{#1.}
\makeatother

\begin{abstract}
\noindent The adjustment of two different selfocs is considered
using both exact formulas for the mode-connection coefficients
expressed in terms of Hermite polynomials of several variables and a
qualitative approach based on the Frank–-Condon principle. Several
examples of the refractive-index dependence are studied and
illustrative plots for these examples are presented. The connection
with the tomographic approach to quantum states of a two-dimensional
oscillator and the Frank–-Condon factors is established.
\end{abstract}


\setcounter{section}{0} \setcounter{equation}{0}

\section{Introduction}

It is known that radiation beams propagating in optical waveguides
can be described by Schr\"{o}dinger-like equations [1--6]; this has
been studied in [7--20]. In this equation, the longitudinal
coordinate $z$ plays the role of time and the radiation wavelength
plays the role of the Planck's constant. The refractive-index
profile is an analog of the potential energy in the quantum
Schr\"{o}dinger equation. The electromagnetic radiation modes in the
optical waveguide are analogs of the wave functions of a quantum
system. This quantumlike picture of the field propagating in the
waveguide was widely used to study the energy distribution among the
modes of successive two (or several) waveguides with different
refractive index profiles, if the distribution in the first
waveguide is known. This problem is equivalent to the
quantum-mechanical problem of the vibronic structure of electronic
lines in the absorption and emission spectra of polyatomic
molecules. In the case of the harmonic potential, the structure is
determined in terms of the Frank-–Condon factor expressed through
Hermite polynomials of several variables [21--23] (see, also
[24--30]). Recently the technique of femtosecond pulses was
developed and applied in different domains of ultrapower laser
physics \cite{Mikheev1, Mikheev2} where the problem of waveguiding
such kinds of radiation also arises \cite{WF1, WF2}.

Our aim here is to consider analogs of the quantum problems of the
molecular spectra discussed in [27--30] to transfer the results of
these investigations into the domain of femtosecond-pulse laser
physics. The other aim is to connect the classical problem of
optical waveguides with the quantum problem of entanglement
\cite{Schrodinger,EPR}. The point is that for states with two
degrees of freedom the notion of entanglement corresponds to the
degree of correlation of observables related to these different
degrees of freedom. In the waveguide picture, an analog of the
entanglement is related to the structure of the modes propagating in
the waveguides. When the initial separable two-mode field in the
first waveguide is propagating into the second waveguide with a
different profile of the refractive index, the entanglement appears
in the arising modes in the second waveguide. The entanglement can
be considered and related to the energy distribution among the field
modes in the second waveguide. Here we discuss this relation and
consider the possibility to find connections with different criteria
of the entanglement [37--41].

The paper is organized as follows.

In Sec. 2, we review the results related to the modes in planar
waveguides with quadratic refractive index. In Sec. 3 the two-mode
problem in selfoc-like waveguides is considered, and an analog of
the two-mode squeezed-light wave function is used for studying the
Frank-–Condon factor. The entanglement analogs of the waveguide
modes are discussed in Sec. 4. The conclusion and perspectives are
presented in Sec. 5.

\section{Planar Waveguides}

In this section, we consider the light-beam propagation from one
planar waveguide with a quadratic refractive index to another
waveguide with the propagation direction along the $z$ coordinate.
The refractive index, being constant along the $y$ and $z$ axes, is
described by a symmetric potential-energy curve of a one-electron
atom as follows:
\begin{equation}n(x)=kx^{2}\; \leftrightarrow\; U(x)=\frac{\omega^{2}x^2}{2}. \end{equation}
The second waveguide has a different dependence of the refractive
index, and its axis is shifted from the axis of the first waveguide
by $d$
\begin{equation}n'(x)=k'(x-d)^{2} \;\leftrightarrow\; U(x)=\frac{\omega'^{2}(x-d)^2}{2}, \end{equation}
which is equivalent to shifting and stretching or shrinking of the
potential-energy curve. Here we take $m=m'=1$.

The most probable energy level to which the electron jumps can be
found in view of the Frank-–Condon principle. Taking into
consideration that the energy levels in a parabolic potential are
distributed as $E_{n}=\hbar\omega(n+1/2)$, the final state can be
easily found.

\begin{figure}[t]
  \begin{minipage}[h]{0.5\linewidth}
        \centering
        \includegraphics[width=250 pt]{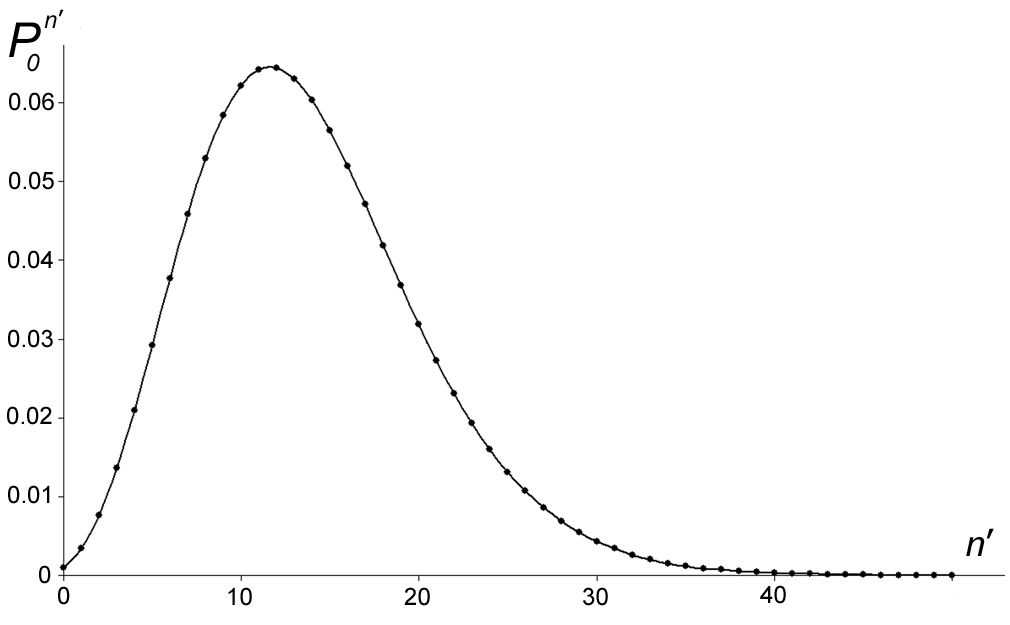}
        \flushleft \hangindent=0.23cm \hangafter=0 \small \textbf{Fig. 1.}
        The plot of $P_{0}^{n'}$ for $\omega'/\omega=3$ and
        $\omega d^{2}=9$.
  \end{minipage}
  \begin{minipage}[h]{0.5\linewidth}
        \centering
\includegraphics[width=250 pt]{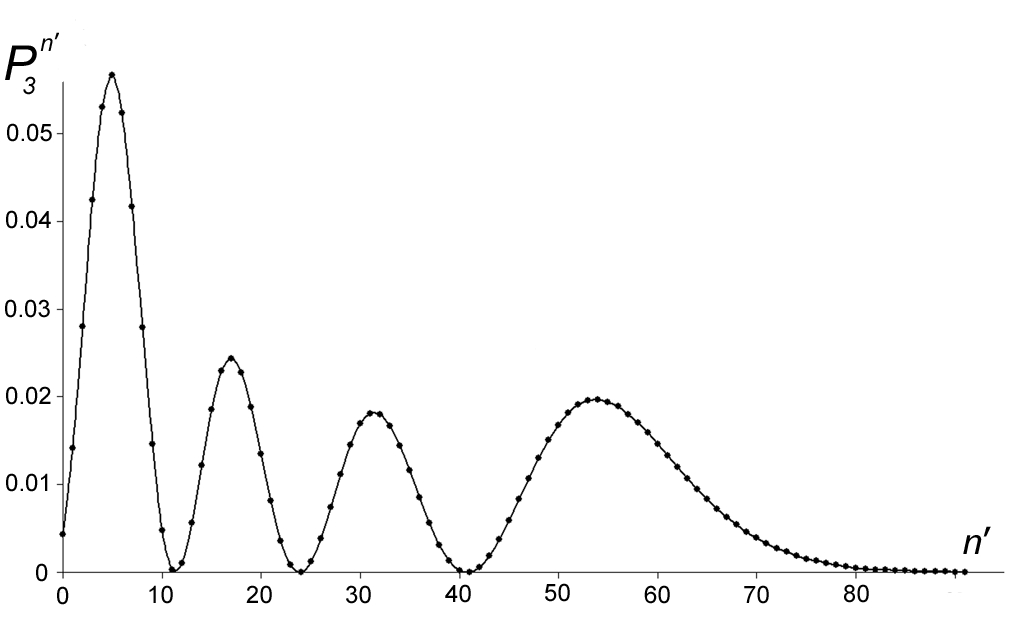}
        \flushleft \hangindent=0.3cm \hangafter=0 \small
        \textbf{Fig.~2.}~The~plot~of~$P_{3}^{n'}$~for~$\omega'/\omega=3$~and~$\omega d^{2}=16$.
  \end{minipage}
\end{figure}

We consider a one-electron atom. The nuclear Hamiltonian in the
Born--Oppenheimer approximation reads
\begin{equation} \mathcal{H}=\frac{1}{2}\hbar \omega \{a,a^{\dagger}\}, \label{Ham}\end{equation}
where $\omega$ is the frequency corresponding to the coordinate $q$,
$a$ and $a^{\dagger}$ are the creation and annihilation operators,
respectively. The wave function of electron in the atom with
Hamiltonian (\ref{Ham}) is described by
\begin{equation} \psi \left(x,n,\omega \right) = \left( \pi^{1/2} 2^{n} n! l(\omega) \right)^{-1/2} \exp \left( -\frac{x^{2}}{2l^{2}(\omega)} \right) H_{n} \left( \frac{x}{l(\omega)} \right), \end{equation}
where $l(\omega)=\sqrt{\hbar/\omega}$, $n$ is a vibrational quantum
number, and $H_{n}(\xi)$ is the $n$th Hermite polynomial. Shifting
the center of the potential from $x=0$ to $x=d$ and changing $\omega
\rightarrow \omega'$ and $n \rightarrow n'$ provides the wave
function
\begin{equation} \psi \left(x,n',\omega',d \right) = \left( \pi^{1/2} 2^{n'} n'! l(\omega') \right)^{-1/2} \exp \left( -\frac{(x-d)^{2}}{2l^{2}(\omega')} \right) H_{n'} \left( \frac{x-d}{l(\omega')} \right). \end{equation}
The overlap integral
\begin{equation} \left< n| n' \right> = \int \psi(x,n,\omega,0) \psi(x,n',\omega',d) \mathbf{d} x \label{nn}\end{equation}
describes the transition probability amplitude for
\begin{equation}\psi(x,n,\omega,0) \rightarrow \psi(x,n',\omega',d). \end{equation}
The integral (\ref{nn}) can be expressed through the Hermite
polynomial of two variables
\begin{equation}\left< n| n' \right> = \left( 2^{n+n'} n!n'! \right) ^{-1/2} \sqrt{\frac{2 l(\omega) l(\omega')}{l(\omega)^{2}+l(\omega')^{2}}} \exp \left( -\frac{d^{2}}{2 \left( l(\omega)^{2}+l(\omega')^{2} \right) } \right) H_{nn'}^{\{R\}} \left(y_{1}, y_{2}
\right).  \end{equation} The arguments of the Hermite polynomial
read
\begin{equation} \begin{array}{l} R=\frac{2}{l(\omega)^{2}+l(\omega')^{2}} \left( \begin{array}{cc}
l(\omega)^{2}-l(\omega')^{2} & -2l(\omega)l(\omega')\\
-2l(\omega)l(\omega') & -l(\omega)^{2}+l(\omega')^{2}\\
\end{array} \right)\\
\left( \begin{array}{c} y_{1} \\ y_{2} \end{array} \right) =
-\frac{dl(\omega)}{l(\omega)^{2}+l(\omega')^{2}} \left(
\begin{array}{c} 1
\\ -l(\omega')/l(\omega)
\end{array} \right)
\end{array}
\end{equation}

The transition probability from the state $\left | n \right >$ to
the state $\left | n' \right >$ is equal to
\begin{equation}P_{n}^{n'} = (\left< n| n' \right>)^{2}. \end{equation}
To illustrate the probability distribution, we present in Fig. 1 a
plot of the function $P_{n}^{n'}$ for the initial level $n=0$,
potential stretching $\omega'/\omega=3$ and shift $\omega d^{2}=9$.
The maximum probability is observed for the thirteenth level. This
result is in agreement with the number calculated directly from the
Frank–-Condon principle. The plot for the transition from $n=3$ with
$\omega'/\omega=3$ and $\omega d^{2}=16$ is shown in Fig. 2. In this
case, the most probable final state is $\left | 5 \right >$.

\section{Elliptic Waveguides}
Now we extend our consideration to the case of elliptical waveguides
with quadratic refractive index, which corresponds to a one-electron
atom with the 2D parabolic potential,
\begin{equation} n(x,y)=k_{x}x^{2}+k_{y}y^{2} \;\leftrightarrow\; U(x,y)=\frac{\omega_{x}x^{2}+\omega_{y}y^{2}}{2}. \label{unshift2d}\end{equation}
In complete analogy with the previous section, the shifted and
deformed potential reads
\begin{equation} U'(x,y)=\frac{\omega'_{x}(x-d_{x})^{2}+\omega'_{y}(y-d_{y})^{2}}{2}. \label{shift2d}\end{equation}
The unshifted wave function of the electron in potential
(\ref{unshift2d}) is
\begin{equation}\psi \left(x,y,n_{x},n_{y},\omega_{x},\omega_{y}, 0,0\right) =  \frac{ \exp \left[ -\frac{1}{2}\left(\frac{x^{2}}{l^{2}(\omega_{x})}+\frac{y^{2}}{l^{2}(\omega_{y})} \right)\right]}{\left( \pi 2^{n_{x}+n_{y}} n_{x}!n_{y}! l(\omega_{x})l(\omega_{y}) \right)^{1/2}} H_{n_{x}} \left( \frac{x}{l(\omega_{x})}\right) H_{n_{y}} \left( \frac{y}{l(\omega_{y})} \right). \end{equation}
In the same way, we can find the probability distribution in the 2D
case for the transition
\begin{equation} \psi \left( x,y,n_{x},n_{y},\omega_{x},\omega_{y}, 0,0 \right) \rightarrow \psi \left( x,y,n'_{x},n'_{y},\omega'_{x},\omega'_{y}, d_{x}, d_{y} \right) \end{equation}
by calculating the overlap integral
\begin{equation} \left< n| n' \right> = \int \psi \left( x,y,n_{x},n_{y},\omega_{x},\omega_{y}, 0,0 \right) \psi \left( x,y,n'_{x},n'_{y},\omega'_{x},\omega'_{y}, d_{x}, d_{y} \right)\, \mathbf{d} x \: \mathbf{d} y. \end{equation}

Figure 3 presents the distribution for the transition from the
ground state for $\omega'_{x}/\omega_{x}=2$,
$\omega'_{y}/\omega_{y}=3$, $\omega_{x}d_{x}^{2}=9$, and
$\omega_{y}d_{y}^{2}=16$. The maximum of this function is observed
for the final state $\left| 23,8 \right>$.

In Fig. 4, a 3D plot of the transition probability for the initial
state $\left| 2,1 \right>$ and $\frac{\omega'_{x}}{\omega_{x}}=2$,
$\frac{\omega'_{y}}{\omega_{y}}=3$, and
$\omega_{x}d_{x}^{2}=\omega_{y}d_{y}^{2}=16$ is presented. The state
$\left| 8,4 \right>$ is the most probable final state.

The pictures are obviously very similar to the planar case. The
transition probability distribution for the initial state
$\left|0,0\right>$ has the Gaussian form. For the transitions from
the excited state, one also can observe a multi-maxima surface.

\begin{figure}[t]
  \begin{minipage}[h]{0.5\linewidth}
        \centering
        \includegraphics[width=255 pt]{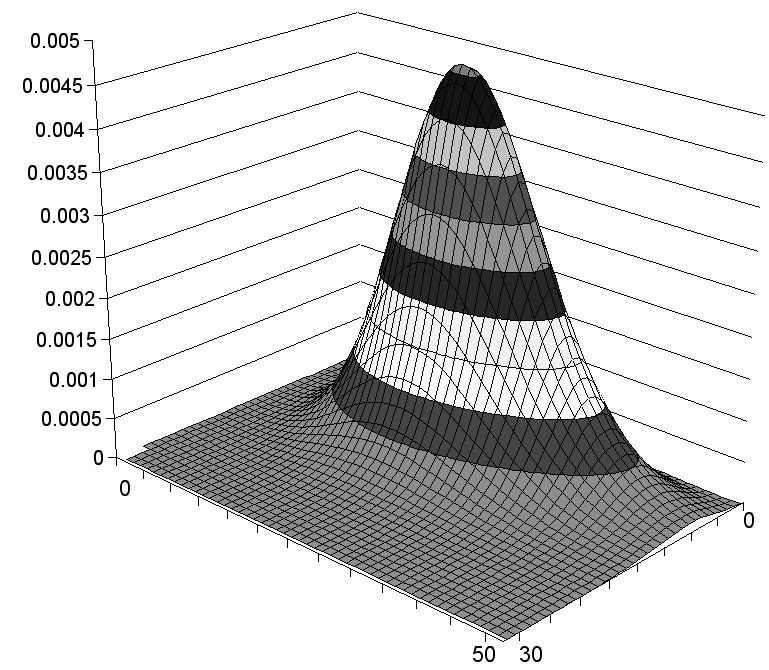}\vspace{5px}
        \flushleft \hangindent=0.23cm \hangafter=0 \small \textbf{Fig.
        3.} Plot of $P_{0,0}^{n'_{x},n'_{y}}$ for $\omega'_{x}/\omega_{x}=2$,
$\omega'_{y}/\omega_{y}=3$, $\omega_{x}d_{x}^{2}=9$, and
$\omega_{y}d_{y}^{2}=16$.
  \end{minipage}
  \begin{minipage}[h]{0.5\linewidth}
        \centering
\includegraphics[width=257 pt]{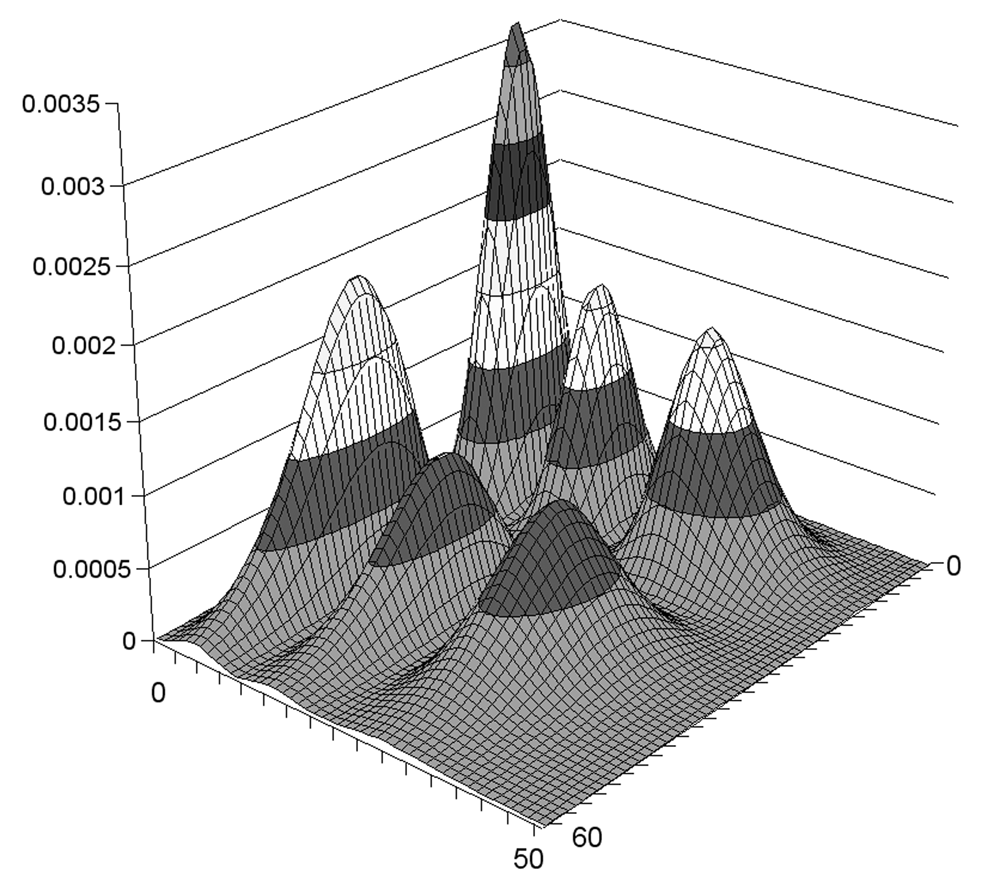}
        \flushleft \hangindent=0.3cm \hangafter=0 \small
        \textbf{Fig. 4.} Plot of $P_{2,1}^{n'_{x},n'_{y}}$ for $\omega'_{x}/\omega_{x}=2$,
$\omega'_{y}/\omega_{y}=3$, $\omega_{x}d_{x}^{2}=16$, and
$\omega_{y}d_{y}^{2}=16$.
  \end{minipage}
\end{figure}

Since there is no $xy$ component in potential (\ref{shift2d}), the
wave created as a result of transition does not contain the
entanglement term. To obtain a state analogous to the entangled
state, one should consider the potential with such a term, i.e.,
\begin{equation} U(x,y)=\frac{\omega_x x^2 + \omega_y y^2 + \gamma x
y}{2}. \end{equation} After the transition, the wave function has
the form
\begin{equation}\psi \left(x,y\right) = \frac{ \exp \left[ -\frac{1}{2}\left(\frac{x^{2}}{l^{2}(\omega_{x})}+\frac{y^{2}}{l^{2}(\omega_{y})} + \gamma x y \right)\right]}{\left( \pi 2^{n_{x}+n_{y}} n_{x}!n_{y}! l(\omega_{x})l(\omega_{y}) \right)^{1/2}} H_{n_{x}} \left(ax+by\right) H_{n_{y}} \left(cx+dy\right). \end{equation}
As a result, the probability distribution is changed slightly in
this case, so we present just plots for potential (\ref{shift2d})
with $\gamma = 0$, keeping in mind that the real plots are very
close to these ones.

\section{Conclusions}
To conclude, we point out the main results of the paper.

We reviewed the approach for describing the light propagation in
optical waveguides using the Fock–-Leontovich approximation. In this
approximation, the equation describing the light intensity and phase
is identical to the Schr\"{o}dinger equation. In view of this,
following \cite{MM8} we applied an analog of the Frank–-Condon
principle to express the energy distribution among the modes in the
optical waveguide by the Frank-–Condon factors. For harmonic
potential, these factors are given in the explicit form in terms of
special functions associated with Hermite polynomials. We made an
analysis of some concrete examples and investigated the
energy-distribution function for the light modes in the optical
waveguide. The distribution functions were shown to have maxima
corresponding to the Frank–-Condon principle.

\section{Acknowledgments}
V.I.M. acknowledges the support of the Russian Foundation for Basic
Research under Projects Nos. 07-02-00598 and 08-02-90300.

\end{document}